# Novel Relations between the Ergodic Capacity and the Average Bit Error Rate[⋆]

Ferkan Yilmaz [1] and Mohamed-Slim Alouini [2]

*Electrical Engineering Program, Division of Physical Sciences and Engineering,*
*King Abdullah University of Science and Technology (KAUST),*
*Thuwal, Mekkah Province, Saudi Arabia.*
[1,2]{ferkan.yilmaz,slim.alouini}@kaust.edu.sa

*Abstract*—Ergodic capacity and average bit error rate have been widely used to compare the performance of different wireless communication systems. As such recent scientific research and studies revealed strong impact of designing and implementing wireless technologies based on these two performance indicators. However and to the best of our knowledge, the direct links between these two performance indicators have not been explicitly proposed in the literature so far. In this paper, we propose novel relations between the ergodic capacity and the average bit error rate of an overall communication system using binary modulation schemes for signaling with a limited bandwidth and operating over generalized fading channels. More specifically, we show that these two performance measures can be represented in terms of each other, without the need to know the exact end-to-end statistical characterization of the communication channel. We validate the correctness and accuracy of our newly proposed relations and illustrated their usefulness by considering some classical examples.

## I. INTRODUCTION

Modern digital communications theory has brought the changes to the radio engineering, one of which is the need for end-to-end performance measurements. The measure of that performance is generally the bit-error rate (BER), which quantifies the reliability of a communication system based on the employed detection scheme, signal-to-noise ratio (SNR), and diversity technique. On the other hand, defining the Shannon capacity [1] of a signaling channel as a maximum error-free rate is the *cornerstone* problem of modern communications theory. Not to mention, indeed, that the Shannon capacity is a more abstract idea than the BER as a performance measure and it leads to other important theoretical conclusions that underlie key concepts of information, entropy, and bandwidth.

[⋆]This work was supported by King Abdullah University of Science and Technology (KAUST).

[⋆]This work has been presented by Ferkan Yilmaz in IEEE International Symposium on Wireless Communication Systems (ISWCS 2011), Aachen, Germany, 6th-9th November, 2011. This work is presented to ensure a date and time stamp from arXiv central authority "http://arxiv.org/" on the purpose of that this material is proven to be authors' scholarly and technical work. This work is moreover presented to share our newly obtained results with the readership in advance of publication under the condition that copyright and all rights therein are retained by authors or by other copyright holders. All persons copying this work and accessing the information inside of it are strictly expected to adhere to the terms, conditions and constraints invoked by each author's copyright. In most cases, this material cannot not be reposted, reannounced, reuploaded and reused without the explicit permission of the authors or other copyright holders.

[⋆]For more information about the related publications, the readers and researchers are referred to "https://sites.google.com/site/ferkanyilmaz/home".

The Shannon capacity of a signaling with bandwidth $W$ over additive white Gaussian noise (AWGN) channels is given in general by $C(\gamma_{end}) \triangleq W \log_2(1 + \gamma_{end})$, where $\gamma_{end}$ is the overall instantaneous conditional SNR (that is, it is the instantaneous SNR at the output of the overall system) and $\log_2(\cdot)$ is the binary logarithm (i.e, it is the logarithm to the base 2). Then, the ergodic capacity, i.e., $C_{avg} \equiv \mathbb{E}[W \log_2(1 + \gamma_{end})]$, where $\mathbb{E}[\cdot]$ the expectation operator, is defined by averaging the instantaneous capacity $C(\gamma_{end})$ over the probability density function (PDF) of the overall instantaneous SNR $\gamma_{end}$ [2, Eq.(15.21)], i.e.,

$$C_{avg} = W \int_0^\infty \log_2(1+\gamma) p_{\gamma_{end}}(\gamma) d\gamma, \qquad (1)$$

where $p_{\gamma_{end}}(\gamma)$ is the probability density (PDF) of the overall instantaneous SNR $\gamma_{end}$. Remembering again that the channel capacity is defined as the error-free maximum rate achievable in the signaling channel, where the definition of *error-free* relies on the fact that there may be a specific signalling technique (for example, it may be the combination of modulation, error correcting codes and linear/non-linear detection techniques) that enables error-free signaling provided that the information rate is below the capacity of the corresponding channel. On the other hand, errors occur even at very low transmission rates. Indeed, noise is ubiquitous, which suggests that some errors are inevitable at any transmission rate. More precisely, assuming that the instantaneous BER over generalized fading channels is compactly expressed by $E(\gamma_{end})$. Then, the average BER $E_{avg} \equiv \mathbb{E}[E(\gamma_{end})]$ is defined by averaging the instantaneous BER $E(\gamma_{end})$ over the PDF of the overall instantaneous SNR $\gamma_{end}$ [2, Section 1.1.3], i.e.,

$$E_{avg} = \int_0^\infty E(\gamma_{end}) p_{\gamma_{end}}(\gamma) d\gamma. \qquad (2)$$

When the signalling channel is fully loaded with the information at the level of channel capacity, the transmission through the channel experiences errors. Regarding this fact and referring to both (1) and (2), we set out to see if there are direct relationships between the ergodic capacity $C_{avg}$ and the average BER $E_{avg}$. As the ergodic capacity $C_{avg}$ increases, the average BER $E_{avg}$ decreases as it seems that they are inversely proportional. Moreover and to the best of our knowledge, the following two important questions have not been answered so far: i)

1) What is the ergodic capacity $C_{avg}$ of a signalling channel if we have only the average BER $E_{avg}$ of a

given specific modulation technique in the corresponding channel?
2) What is the average BER $E_{avg}$ of a given specific modulation technique if we have only the ergodic capacity $C_{avg}$ of the corresponding channel?

Viewed from a somewhat broader perspective, each question above is admittedly opening a set of new ideas and techniques in information theory, communications theory, or coding theory, although all of these questions have been defined with different scopes as well.

In this paper, we investigate these two questions given above and we provide two new simple analytical relations linking the ergodic capacity $C_{avg}$ of a signalling channel and the average BER $E_{avg}$ of a specific modulation scheme in the corresponding channel. Additionally, three possible examples of these analytical tools are demonstrated in this paper in order to show that these two techniques are promising technique regarding the performance measures of digital communication systems. More precisely, *without the need to know the statistical characterizations (such as PDF, cumulative distribution function (CDF), moment generating function (MGF), moments, Laguerre moments, etc.) for the overall instantaneous SNR of the communication system.*, one can readily obtain the ergodic capacity $C_{avg}$ inn terms of the empirically obtained average BER $E_{avg}$, and vice versa the average BER $E_{avg}$ in terms of the ergodic capacity $C_{avg}$.

The remainder of this paper is organized as follows. Section II is devoted to the mathematical development of concepts related to the two questions mentioned above. After defining the certainty measure, the relation between the average BER, average certainty and the outage probability are underlined by the PDF of the certainty measure. In particular, we show in Section III that, for communication systems signalling with binary modulation schemes over generalized fading channels, the ergodic capacity $C_{avg}$ can be obtained in terms of the average BER $E_{avg}$. In Section IV, we show on the other hand that the average BER $E_{avg}$ can be computed from the ergodic capacity $C_{avg}$ without the need to know the exact the statistical characterization of the overall end-to-end instantaneous SNR $\gamma_{end}$. It should be mentioned that several proofs of our results have intentionally been omitted due to space limitation but are available in the extended version of this paper. In addition, all these results have been checked numerically for their correctness and accuracy.

## II. BACKGROUND RELATIONS

The present section is devoted to the mathematical development of the basic and fundamental concepts related to the two key questions mentioned in the previous section. We shall develop our main definitions and theorems in full generality, for an arbitrary communication system signalling over generalized fading channels, and we shall use the language of probability theory throughout this paper to facilitate the mathematical exposition of the relation between the ergodic capacity $C_{avg}$ of a signalling channel and the average BER $E_{avg}$ of a specific modulation scheme in the corresponding channel.

Let us consider the instantaneous BER $E(\gamma_{end})$ of a specific communication system, where $\gamma_{end} \in \mathbb{R}^+$ is the instantaneous SNR as mentioned before. When for a specific value $\gamma_{end}$, the instantaneous BER $E(\gamma_{end}) = 1/2$, then the signalling channel becomes a fully dissipated channel, which means that, referring to the channel certainty, the transferred entropy through signalling channel becomes zero. From this point of view, one can conclude that there exists a kind of certainty measure, that is

$$\Phi(\gamma_{end}) = 1 - 2E(\gamma_{end}), \qquad (3)$$

which possesses the information about the channel certainty, and it has a close relation with the channel capacity since the channel capacity is the maximum transmission rate without a transmission error as mentioned before. In more details, the signaling channel is fully dissipated / lossy with $\Phi(\gamma_{end}) = 0$ or it is error-free with $\Phi(\gamma_{end}) = 1$. It is important to note that, specifically from a practical point of view, the instantaneous BER $E(\gamma_{end})$ has always a distribution between zero and half for all binary modulation schemes (such as binary phase shift keying (BPSK), differentially encoded BPSK (BDPSK), binary frequency shift keying (BFSK) and non-coherent BFSK (NC-FSK) binary schemes). As a result, the certainty measure $\Phi(\gamma_{end})$ is distributed between zero and one as expected.

With the aforementioned deduction conditioned on that the knowledge of the instantaneous SNR $\gamma_{end}$ is precisely known, the certainty measure $\Phi(\gamma_{end})$ can be readily defined as $\Phi(\gamma_{end}) = \mathbb{E}_\phi[\theta(\gamma_{end} - \phi)]$, where $\mathbb{E}_\psi[\cdot]$ is the expectation operator with respect to the random variable $\psi$, and $\theta(\cdot)$ is the Heaviside's theta (step) function [3, Eq.(1.8.3)], and where $\phi$ is the instantaneous certainty measure having a non-negative distribution following the PDF $p_\phi(\varphi) = \mathbb{E}_\phi[\delta(\varphi - \phi)]$, where $\delta(\cdot)$ is the Dirac's delta function [3, Eq.(1.8.1)], that is readily defined as

$$p_\phi(\varphi) = \frac{\partial \Phi(\varphi)}{\partial \varphi}, \quad \varphi \in \mathbb{R}^+, \qquad (4)$$

with the intuitive observation that $\Phi(\gamma_{end}) = \int_0^{\gamma_{end}} p_\phi(\varphi) d\varphi$. Accordingly, both (3) and (4) suggest that the signalling error occurred through end-to-end transmission can be, without loss of generality, modeled as non-negative random variable (RV). Then, referring to (2), the average BER $E_{avg} = \mathbb{E}[E(\gamma_{end})]$ of an arbitrary communication system signalling over generalized fading channels can be attained from the outage probability $P_{\gamma_{end}}(\varphi)$ as shown in the following theorem.

**Theorem 1.** *The average BER $E_{avg} = \mathbb{E}[E(\gamma_{end})]$ over generalized fading channels is alternatively given by*

$$E_{avg}(\bar{\gamma}_{end}) = \frac{1}{2} - \frac{1}{2}\int_0^\infty \widehat{P}_{\gamma_{end}}(\varphi) p_\phi(\varphi) d\varphi, \qquad (5)$$

*where $\bar{\gamma}_{end}$ is the average SNR defined as $\bar{\gamma}_{end} = \mathbb{E}[\gamma_{end}]$, and $\widehat{P}_{\gamma_{end}}(\gamma) \equiv \mathbb{E}[\theta(\gamma_{end} - \gamma)] = \int_\gamma^\infty p_{\gamma_{end}}(u) du$ is the complementary outage probability that the instantaneous SNR $\gamma_{end}$ exceeds a certain specified threshold $\gamma$.*

*Proof:* The proof is omitted due to space limitation. ∎

It may be here useful to define the average certainty measure $\Phi_{avg} \equiv \mathbb{E}[\Phi(\gamma_{end})]$. Referring to both (2) and (3), the averaged certainty measure $\Phi_{avg}$ can obtained as $\Phi_{avg} = 1 - 2E_{avg}$. On the other hand, utilizing Theorem 1, the average certainty $\Phi_{avg}$ can be alternatively given in the following corollary.

**Corollary 1.** *The average certainty $\Phi_{avg} = \mathbb{E}[\Phi(\gamma_{end})]$ over generalized fading channels is alternatively given by*

$$\Phi_{avg}(\bar{\gamma}_{end}) = \int_0^\infty \widehat{P}_{\gamma_{end}}(\varphi) p_\phi(\varphi) d\varphi. \quad (6)$$

*Proof:* The proof is obvious. ∎

As an example, a unified instantaneous BER expression for binary modulation schemes is known to be given in [4, Eq.(13)] by Wojnar as

$$E(\gamma_{end}) \triangleq \frac{\Gamma(b, a\gamma_{end})}{2\Gamma(b)}, \quad (7)$$

where $a$ depends on type of modulation scheme ($\frac{1}{2}$ for orthogonal FSK and $1$ for antipodal PSK), $b$ depends on type of detection technique ($\frac{1}{2}$ for coherent and $1$ for non-coherent), and $\Gamma(\cdot)$ and $\Gamma(\cdot,\cdot)$ are the Gamma function [5, Eq. (6.1.1)] and the complementary incomplete Gamma function [5, Eq. (6.5.3)], respectively. Particularly in connection with (7), applying [6, Eqs. (8.4.16/1) and (8.4.16/1)] on (7), we are now able to write the certainty measure $\Phi(\gamma_{end})$ for binary modulation schemes as

$$\Phi(\gamma_{end}) = \frac{1}{\Gamma(b)} G_{1,2}^{2,0}\left[a\gamma_{end} \,\middle|\, \begin{matrix} 1 \\ b, 0 \end{matrix}\right], \quad (8)$$

where $G_{p,q}^{m,n}[\cdot]$ is the Meijer's G function [6, Eq. (8.3.22)]. Accordingly using (4), (7) and (8), the PDF of the instantaneous certainty measure $\phi$, i.e., $p_\phi(\varphi)$ for binary modulation schemes is obtained as

$$p_\phi(\varphi) = \frac{(b/a)^b}{\Gamma(b)} \varphi^{b-1} e^{-(b/a)\varphi}, \; \varphi \in \mathbb{R}^+, \quad (9)$$

which is the Gamma distribution with the parameters $\mathbb{E}[\phi] = a$ and $\mathbb{E}[\phi^2]/\mathbb{E}[\phi]^2 = (b+1)/b$. In particular, substituting $b = 1$ in (9) results in the exponential PDF, i.e., $p_\phi(\varphi) = (1/a)\exp(-\varphi/a)$. In addition, substituting $b = 1/2$ reduces (9) to the power distribution of the half-Gaussian RV, i.e., $p_\phi(\varphi) = \exp(-\varphi/(2a))/\sqrt{2\pi a \varphi}$. By means of using (2), it is at this point convenient to give the averaged certainty $\Phi_{avg}$ for binary modulation schemes as

$$\Phi_{avg}(\bar{\gamma}_{end}) = \frac{(b/a)^b}{\Gamma(b)} \int_0^\infty \widehat{P}_{\gamma_{end}}(\varphi) \varphi^{b-1} e^{-(b/a)\varphi} d\varphi. \quad (10)$$

It is useful to mention that we have described in this section the relations between the average BER $E_{avg}(\bar{\gamma}_{end})$, the average certainty $\Phi_{avg}(\bar{\gamma}_{end})$ and the complementary outage probability $\widehat{P}_{\gamma_{end}}(\gamma)$, which are entirely easy-to-obtain (i.e., without need to have the exact PDF $p_{\gamma_{end}}(\gamma)$ of the overall instantaneous SNR $\gamma_{end}$ which is harder to obtain experimentally than that the outage probability) and obviously easy-to-use.

In the following sections, in addition to better illustrate the usefulness of the relations mentioned above, we shall obtain direct relations between the ergodic capacity $C_{avg}$ and the average BER $E_{avg}$ for communication systems signalling with binary modulation schemes over generalized fading channels.

## III. Ergodic Capacity Obtained Through the Bit-Error Rate of Binary Modulations

We show in this section that, for wireless communication systems signalling with binary modulation schemes over generalized fading channels, the ergodic capacity $C_{avg}$ can be implicitly obtained in terms of the average BER $E_{avg}$ by using the results given in (8) and (10). This relation is indeed a promising technique for quality / performance assessment of a wireless communication systems due to the fact that the average BER $E_{avg}$ is more measurable and controllable than the ergodic capacity $C_{avg}$.

**Theorem 2.** *Consider a communication system signalling with bandwidth $W$ over generalized fading channels. Then, without having the exact the statistical characterization of the overall instantaneous SNR $\gamma_{end}$, the ergodic capacity $C_{avg}$ can be obtained directly from the average BER $E_{avg}$ for binary modulation schemes as*

$$C_{avg}(\bar{\gamma}_{end}) = \int_0^\infty \mathcal{Z}_{a,b}(u) \left[1 - 2E_{avg}(u\bar{\gamma}_{end})\right] du, \quad (11)$$

*where the information certainty content $\mathcal{Z}_{a,b}(u)$ is given by*

$$\mathcal{Z}_{a,b}(u) = \frac{W\Gamma(b)}{\log(2)u} G_{1,2}^{1,1}\left[au \,\middle|\, \begin{matrix} 0 \\ 0, 1-b \end{matrix}\right], \quad (12)$$

*where $a$ and $b$ are the modulation specific parameters defined before, and where $\log(\cdot)$ is the natural logarithm.*

*Proof:* The proof is omitted due to space limitation. ∎

It is worthy to mention that Theorem 2 is quite useful for the analysis of wireless communication systems. Specifically, in order to find the ergodic capacity $C_{avg}$, the proposed technique in Theorem 2 eliminates the necessity to have the knowledge of the communication medium. More precisely, a key indication of Theorem 2 is that there is no need to know the details of the transmission medium and the statistical characterizations of the overall instantaneous SNR $\gamma_{end}$. One just needs to know i) the binary modulation scheme employed for signalling, ii) the signalling bandwidth $W$ and iii) the average BER $E_{avg}$ curve over the average SNR $0 < \bar{\gamma}_{end} < \infty$, in order to obtain the ergodic capacity.

**Corollary 2.** *The information certainty content $\Phi_{a,b}(u)$ given by (12) can also be represented as*

$$\mathcal{Z}_{a,b}(u) = -\frac{W(b-1)\gamma(b-1, -au)}{\log(2)a^{b-1}u^b \exp(au + i\pi b)}, \quad (13)$$

*where $i = \sqrt{-1}$ is the unit imaginary number, and $\gamma(\cdot,\cdot)$ is the complementary incomplete Gamma function [5, Eq. (6.5.2)].*

*Proof:* The proof is omitted due to space limitation. ∎

Note that even if the novel integral relation given by Theorem 2, defining the relation between the ergodic capacity $C_{avg}$ and the average BER $E_{avg}$, is simple to use, we can specifically get its finite ($N$-terms) sum approximation converging rapidly and steadily and requiring few terms for an accurate result, as shown in the following corollary.

**Corollary 3.** *The novel relation between the ergodic capacity $C_{avg}$ and the average BER $E_{avg}$, i.e., Theorem 2 can be given*

in a finite ($N$-terms) sum approximation for binary modulation schemes as

$$C_{avg}(\bar{\gamma}_{end}) = \sum_{n=1}^{N} w_n \mathcal{Z}_{a,b}(s_n) \left[1 - 2E_{avg}(s_n \bar{\gamma}_{end})\right], \quad (14)$$

where both $w_n$ and $s_n$ are defined in [7, Eqs. (22) and (23)], respectively.

*Proof:* By changing the variable of the integration in (11) as $u = \tan(s)$ and then using the Gauss-Chebyshev quadrature (GCQ) formula [5, Eq.(25.4.39)], the proof is obvious. ■

Although the novel relation presented in Theorem 2 is simple and straightforward to use, let us consider some of its example applications in order to have a broad conceptual understanding of the relation and then to double check its analytical correctness.

**Example 1** (Average BER to Ergodic Capacity in Closed–Form). Let us consider the performance measures of binary modulation over Rayleigh fading channels as a simple example, whose well-known unified average BER $E_{avg}$ is given for all binary modulation schemes in [8, Eq. (16)], as

$$E_{avg}(\bar{\gamma}_{end}) = \frac{1}{2} - \frac{1}{2}\left(\frac{a\bar{\gamma}_{end}}{1 + a\bar{\gamma}_{end}}\right)^b, \quad (15)$$

where $a$ and $b$ are the modulation parameters as explained before. Substituting both (13) and (15) into (11) and then using [5, Eq. (6.5.36)], we have

$$C_{avg}(\bar{\gamma}_{end}) = \frac{W}{\log(2)} \int_0^\infty \frac{1}{u} G_{1,2}^{1,1}\left[au \;\middle|\; \begin{matrix} 0 \\ 0, 1-b \end{matrix}\right] \times G_{1,1}^{1,1}\left[au\bar{\gamma}_{end} \;\middle|\; \begin{matrix} 1 \\ b \end{matrix}\right] du. \quad (16)$$

Performing some simple algebraic manipulations using [6, Eqs. (2.24.2/2) and (8.4.16/14)] and then utilizing [5, Eq. (6.5.15)], the ergodic capacity $C_{avg}$ of the Rayleigh fading channels is readily obtained as

$$C_{avg}(\bar{\gamma}_{end}) = \frac{W}{\log(2)} \exp\left(\frac{1}{\bar{\gamma}_{end}}\right) E_1\left(\frac{1}{\bar{\gamma}_{end}}\right), \quad (17)$$

which is in perfect agreement with [9, Eq. (34)], where $E_n(\cdot)$ is the exponential integral [5, Eq.(5.1.4)]. □

In short, this example shows that the ergodic capacity $C_{avg}$ of a communication systems can be obtained from the average BER $E_{avg}$ of a specific binary modulation scheme in the same channel. In contrast, the next example is associated with the implementation and design of a communication system.

**Example 2** (Average BER to Ergodic Capacity Using Numerical Techniques). At the implementation and design stage, the average BER $E_{avg}$ can often be experimentally measured for a specific binary modulations for an enough number of SNR values. It is assumed that we have a set of distinct SNR values such that $0 \leq \bar{\gamma}_1 \leq \bar{\gamma}_2 \leq \ldots \leq \bar{\gamma}_N$, where $\bar{\gamma}_{min} < 10\log_{10}(\bar{\gamma}_n) < \bar{\gamma}_{max}$ for all $n \in \{1, 2, .., N\}$. Furthermore, $\bar{\gamma}_{dB}$ has to be experimentally chosen as small as possible since the averaged certainty measure $\Phi_{avg}$ goes to zero with the low SNR values (i.e., the average BER $E_{avg}$ performance at the low-SNR region determines the ergodic capacity $\Phi_{avg}$ of the channel).

Then, assume that the average BER $E_{avg}$ of the corresponding system for these SNR values are experimentally measured as $E_1, E_2, \ldots, E_N$. Eventually, using available interpolation and extrapolation techniques [10], [11], the average BER $E_{avg}$ can be readily and accurately approximated as

$$E_{avg}(\bar{\gamma}_{end}) \approx \mathcal{E}_{int}\left(\bar{\gamma}_{end}; \{\bar{\gamma}_n\}_1^N, \{E_n\}_1^N\right) \quad (18)$$

where the interpolation/extrapolation technique $\mathcal{E}_{int}(\cdot;\cdot,\cdot)$ are implemented as a built-in function in standard mathematical software packages such as MATHEMATICA®. As a result and using (18) in (11), the ergodic capacity $C_{avg}$ can be approximated as

$$C_{avg}(\bar{\gamma}_{end}) \approx \lim_{\bar{\gamma}_{dB} \to \infty} \int_{10^{-\bar{\gamma}_{dB}/10}}^{10^{\bar{\gamma}_{dB}/10}} \mathcal{Z}_{a,b}(u) \times \left[1 - 2\mathcal{E}_{int}\left(u\bar{\gamma}_{end}; \{\bar{\gamma}_n\}_1^N, \{E_n\}_1^N\right)\right] du. \quad (19)$$

As an illustration of this example, consider the BPSK performance curves of an $L$-branch diversity receiver signalling over assumably an unknown generalized fading channels with bandwidth $W$ are depicted in Fig. 1 for different values of an implementation parameter (i.e., it is the parameter that directly or obliquely effects on the distribution of overall instantaneous SNR $\gamma_{end}$). Without having information about neither the type of the $L$-branch diversity receiver nor the statistical properties of the wireless fading channel that the communication system is subjected to, the corresponding ergodic capacity $C_{avg}$ can be readily computed by means of (19), yielding results in perfect agreement with the correct analytical results as illustrated in details in Fig. 2 at the top of the next page. □

### IV. BIT-ERROR RATE OF THE BINARY MODULATION SCHEMES OBTAINED THROUGH THE ERGODIC CAPACITY

The error caused by the transmission of information through a communication channel is clearly connected to the uncertainty of the corresponding communication system. We show in this subsection that, for communication systems signalling with binary modulation schemes over generalized fading channels, the average BER $E_{avg}$ can also be implicitly given in terms of the ergodic capacity $C_{avg}$ using the results given in (8) and (10).

**Theorem 3.** *Consider a communication system signalling with bandwidth $W$ over generalized fading channels. Then, without the need to have the exact statistical characterization of the overall instantaneous SNR $\gamma_{end}$, the average BER $E_{avg}$ for binary modulation schemes can be obtained directly from the ergodic capacity $C_{avg}$ as follows*

$$E_{avg}(\bar{\gamma}_{end}) = \frac{1}{2} - \int_0^\infty \widehat{\mathcal{Z}}_{a,b}(u) \Im\left\{C_{avg}(-u\bar{\gamma}_{end})\right\} du, \quad (20)$$

*where $\Im\{\cdot\}$ gives the imaginary part of its argument and where the information certainty content $\widehat{\mathcal{Z}}_{a,b}(u)$ is given by*

$$\widehat{\mathcal{Z}}_{a,b}(u) = \frac{\log(2)}{2\pi a W \Gamma(b)} \left(\frac{a}{u}\right)^{b+1} \exp\left(-\frac{a}{u}\right), \quad (21)$$

*where $a$ and $b$ are the parameters of the binary modulation scheme as explained before.*

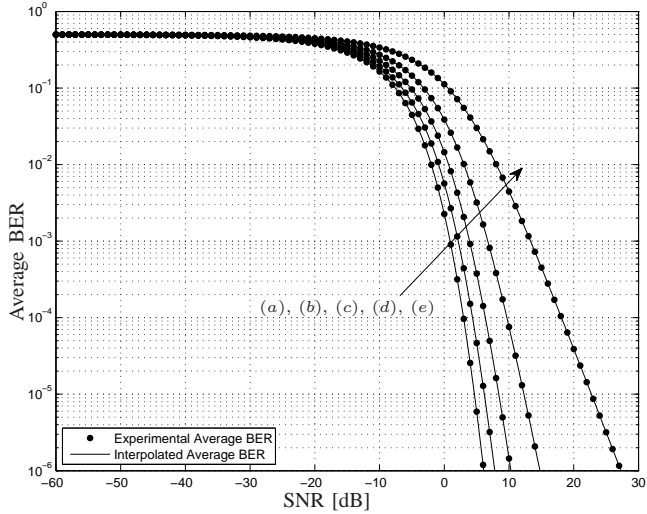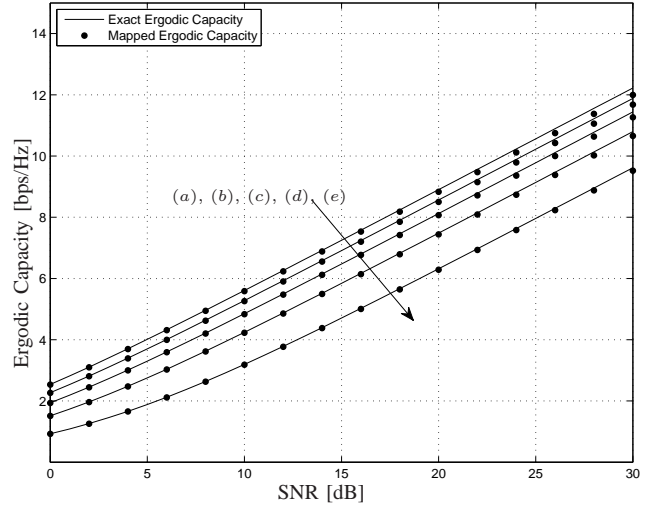

Fig. 1. Average BER $E_{avg}$ of the $L$-branch diversity receiver over assumably unknown generalized fading channels.

Fig. 2. Ergodic Capacity $C_{avg}$ of the $L$-branch diversity receiver over assumably unknown generalized fading channels.

*Proof:* The proof is omitted due to space limitation. ∎

It is obvious that the numerical results and applications illustrated by Example 1 and Example 2 can also be readily used with Theorem 3. So let us consider now the Nakagami-$m$ fading channels as a different example, in order to validate the correctness and numerical accuracy of Theorem 3.

**Example 3** (Ergodic Capacity to Average BER in Closed–Form)**.** The ergodic capacity $C_{avg}$ with bandwidth $W$ over Nakagami-$m$ fading channels is given in [12, Eq. (33)] by

$$C_{avg}(\bar{\gamma}_{end}) = \frac{W}{\log(2)\Gamma(m)} G_{2,3}^{3,1}\left[\frac{m}{\bar{\gamma}_{end}} \,\middle|\, \begin{array}{c} 0,1 \\ m,0,0 \end{array}\right], \quad (22)$$

where the parameter $m$ denotes the fading figure (i.e., diversity order) which ranges from $1/2$ to the infinity [2, Section 2.2.1.4]. After performing some algebraic manipulations, we have

$$\Im\left\{C_{avg}(-\bar{\gamma}_{end})\right\} = \frac{\pi W \Gamma(m, m/\bar{\gamma}_{end})}{\log(2)\Gamma(m)}. \quad (23)$$

Then, substituting (23) into (20) yields

$$E_{avg}(\bar{\gamma}_{end}) = \frac{1}{2} - \frac{a^b}{2\Gamma(b)\Gamma(m)} \times$$
$$\int_0^\infty u^{b-1} \exp(-a\,u)\,\Gamma(m, u\,m/\bar{\gamma}_{end})\,du. \quad (24)$$

Finally, using [13, Eq. (2.10.3/2)], (24) reduces as expected to the well-known generalized average BER over Nakagami-$m$ fading channels [8, Eq. (15)], i.e.,

$$E_{avg}(\bar{\gamma}_{end}) = \frac{\Gamma(m+b)}{2\Gamma(b)\Gamma(m+1)}\left(\frac{a\bar{\gamma}}{m+a\bar{\gamma}}\right)^b \times$$
$$\left(\frac{m}{m+a\bar{\gamma}}\right)^m {}_2F_1\left[1, m+b; m+1; \frac{m}{m+a\bar{\gamma}}\right], \quad (25)$$

where ${}_2F_1[\cdot;\cdot;\cdot]$ is the well-known Gaussian hypergeometric function [6, Eq. (7.2.1.1)]. □

As a result, the mapping proposed in Theorem 3 is a promising technique for directly obtaining the average BER performance of digital communications systems.

## V. CONCLUSION

In this paper, we define the channel certainty measure and then propose novel relations between the ergodic capacity (Shannon capacity) and the average BER performance of an overall communication system using binary modulation schemes for signalling with bandwidth $W$ and operating over generalized fading channels. More specifically, these relations suggest that, even without the need to know the exact end-to-end statistical characterization of the overall communication system, the ergodic capacity can be obtained through the medium of its corresponding average BER, and the average BER can also be explicitly obtained through the ergodic capacity. We finally validated the correctness and accuracy of our newly proposed relations and illustrated their usefulness by considering a few classical examples.